\documentclass[12pt, letterpaper]{article}

\usepackage{amssymb,amsmath,amsfonts,eurosym,geometry,ulem,graphicx,caption,color,setspace,sectsty,comment,footmisc,pdflscape,array, natbib, svg, booktabs}
\usepackage{siunitx} 
\usepackage{threeparttable} 
\usepackage{float}
\usepackage{placeins}
\usepackage{graphicx}
\usepackage[dvipsnames]{xcolor} 

\usepackage[pdftex,
            pdfauthor={Hyunso Kim; Hyo Kang; Jaeyong Song},
            pdftitle={Generative AI Fuels Solo Entrepreneurship, but Teams Still Lead at the Top},
            pdfkeywords={Generative AI; Artificial Intelligence; Entrepreneurial Entry; Experimentation; Venture Creation; Venture Quantity; Venture Quality; Solo Entrepreneurship; Solopreneurship; Team Advantage; Founding Teams; Technological Change; Product Hunt}]{hyperref}

\usepackage{microtype,subcaption,ragged2e}

\newcolumntype{L}[1]{>{\raggedright\let\newline\\arraybackslash\hspace{0pt}}m{#1}}
\newcolumntype{C}[1]{>{\centering\let\newline\\arraybackslash\hspace{0pt}}m{#1}}
\newcolumntype{R}[1]{>{\raggedleft\let\newline\\arraybackslash\hspace{0pt}}m{#1}}

\hypersetup{
    colorlinks=true,
    allcolors=MidnightBlue 
    }

\title{Generative AI Fuels Solo Entrepreneurship, but Teams Still Lead at the Top}

\author{Hyunso Kim\thanks{ SNU Business School, Seoul National University. \texttt{hyunso.kim@snu.ac.kr}}
\and
Hyo Kang\thanks{ SNU Business School, Seoul National University. Co-corresponding author. \texttt{hyokang@snu.ac.kr} }
\and
Jaeyong Song\thanks{SNU Business School, Seoul National University. Co-corresponding author. \texttt{jsong@snu.ac.kr} }
}

\date{\today}

\begin{document}

\maketitle

\begin{abstract}
\noindent Recent advances in generative artificial intelligence (AI) are reshaping who enters entrepreneurship, but not who reaches the top of the quality distribution. Using data on over 160,000 product launches on Product Hunt, we find that entrepreneurial entry increased sharply following the public release of ChatGPT-3.5, driven disproportionately by solo entrepreneurs. This shift toward solo entry is particularly pronounced in categories that historically favored team-based ventures. However, much of this growth reflects low-commitment, experimental entry and does not translate into greater representation among the highest-quality outcomes. Team-based ventures are increasingly dominant in the top tiers of platform rankings. These findings suggest that generative AI lowers barriers to solo entrepreneurship while reinforcing team-based advantages.
\vspace{0.75em}

\end{abstract}

\newpage

\doublespacing
\pagenumbering{arabic}

\section{Introduction}
Recent advances in generative artificial intelligence (AI) represent a major technological inflection point with the potential to reshape economic activity across a wide range of sectors. Following the public release of large language models such as ChatGPT-3.5 in November 2022, hundreds of millions of technical and non-technical users worldwide gained access to tools capable of generating code, designing interfaces, producing content, and iterating on product ideas at unprecedented speed and low cost \citep{brynjolfsson2025generative, epstein2023art, peng2023impact, bick2024rapid}. These developments raise a fundamental question for entrepreneurship: does generative AI change who enters entrepreneurship, and what are the implications for the quality of new ventures?

Literature in innovation economics and entrepreneurship emphasizes that the contribution of entrepreneurship to economic growth depends not on the overall rate of venture creation, but on a small subset of high-growth, innovation-driven ventures \citep{guzman2020state, botelho2026innovation}. This raises an important question during periods of technological change. Although new technologies may lower barriers to entry and increase the number of new ventures, it is less clear whether such expansion translates into higher-quality outcomes.

Recent developments in generative AI make this question especially salient. Early evidence suggests that generative AI lowers entrepreneurial entry barriers and enables leaner organizational forms, allowing individuals to perform tasks that previously required broader founding teams \citep{cai2025aico}. At the same time, a recent field experiment within a large corporation shows that although AI can substantially improve individual performance, teams may be better positioned to capture its benefits \citep{dellacqua2025cybernetic}. These findings together point to competing possibilities. Generative AI may democratize entrepreneurship by enabling solo founders, yet teams may retain advantages in transforming opportunities into superior outcomes. However, existing evidence remains fragmented, focusing either on entry patterns without outcomes or on task-level performance in non-entrepreneurial organizational settings.

We tackle this question by examining a novel dataset of more than 160,000 entrepreneurial entries on Product Hunt, an online platform where entrepreneurs introduce new digital products to the public. A key advantage of this setting is that it allows us to jointly observe both entrepreneurial entry and immediate market-based evaluations of the entries, enabling a direct assessment of how changes in entry translate into differences in early quality outcomes. Although not all such entries correspond to incorporated firms, and some represent small or short-lived initiatives, this setting captures early-stage market exposure before many ventures exit or attract external funding. In doing so, we move beyond conventional measures such as firm registration or venture capital funding, which typically capture only a selective subset of ventures after they have survived to later stages and taken on more evolved organizational structures.

We leverage the public release of ChatGPT-3.5 in November 2022 as a quasi-experimental setting in which potential entrepreneurs gained access to generative AI to help create new products. Using a difference-in-differences and event study framework, we first examine differential changes in entrepreneurial entry between solo and team-based ventures before and after the release. We then explore how these changes vary across categories, examining whether the impact of generative AI is concentrated or broadly distributed. Finally, we examine how venture quality evolves following the shock, using post-launch activity and platform-based rankings to assess overall outcomes and variation across the quality distribution.

Our findings reveal a growing and remarkable divergence between the quantity and quality of entrepreneurship following the emergence of generative AI. We observe a sharp expansion in entry driven predominantly by solo entrepreneurs, with the solo share increasing most in categories that historically relied on team-based ventures (e.g., Software-as-a-service and Fintech). However, much of this expansion reflects a shift toward low-commitment, short-lived entry and the introduction of more distinct combinations of product attributes, suggesting increased experimentation among solo entrepreneurs. Early market evaluations indicate that team-based ventures continue to account for a large share of the highest-ranked outcomes, with their share increasing further after the introduction of generative AI. Put differently, generative AI affected entrepreneurship unevenly, enabling solo entrepreneurs to experiment with novel ideas more easily while allowing teams to further enhance the quality of their products. 

This pattern suggests that, rather than eliminating constraints, generative AI induces a broader shift in where constraints arise in entrepreneurship. As barriers to entry fall, bottlenecks move from initial venture creation toward the development, coordination, and refinement required for successful outcomes.

\section{Theoretical Background}

This paper contributes to several strands of economics and entrepreneurship literature. First, we contribute to the economics of innovation and entrepreneurship research by providing one of the first large-scale empirical analyses of how generative AI reshapes both entrepreneurial entry and early-stage venture outcomes. Prior work emphasizes that economic growth is driven by a small subset of high-growth, innovation-driven ventures \citep{guzman2020state, botelho2026innovation}. We show how generative AI reshapes the composition around this subset by expanding entry through a shift toward solo-founded ventures, while high-quality outcomes remain concentrated among team-based ventures. In doing so, we present evidence that complement emerging findings that generative AI lowers entry barriers and enables leaner ventures \citep{cai2025aico}, while team-based advantages persist even with AI \citep{dellacqua2025cybernetic}. By jointly examining who enters (solo versus team) and who wins in an entrepreneurial setting, we show that generative AI broadens entrepreneurial entry without proportionally increasing the incidence of high-quality, growth-oriented entrepreneurship, suggesting a growing divergence of quantity and quality of new ventures in the generative AI era. 

Second, we contribute to the literature on entrepreneurial strategy and experimentation by showing how a general-purpose technology like generative AI reshapes who engages in experimentation, how it is conducted, and the resulting impact on outcome distributions. Prior work shows that existing digital tools may lower the cost of testing business ideas and improve performance but their uneven adoption limits our understanding of how these effects generalize across ventures \citep{koning2022experiment}. At the same time, related work emphasizes that entrepreneurs often face multiple ex ante equally viable strategic alternatives such that analysis alone does not yield a unique optimum and experimentation becomes a necessary mechanism for resolving uncertainty \citep{gans2019foundations}. We extend these perspectives by showing that the ability to act on multiple viable strategies is not uniform across entrepreneurs and depends on the cost and accessibility of technology that enables experimentation. When experimentation becomes broadly accessible through the introduction of generative AI, it expands predominantly among solo entrepreneurs. This expansion increasingly takes the form of one-off product launches and broader recombination of ideas which leads to distinct products. We further reveal that this shift redistributes gains toward intermediate outcomes rather than the upper tail without altering which ventures reach the highest levels of success. This highlights a disconnect between the expansion of experimentation and the concentration of top outcomes driven by systematic differences between solo and team-founded ventures.

Third, we contribute to the emerging literature on the economic impact of generative AI by linking micro-level productivity effects to early-stage entrepreneurial outcomes. Recent work shows that generative AI can improve individual productivity particularly among less experienced or lower-skilled workers \citep{brynjolfsson2025generative}. At the same time, macroeconomic analyses suggest that aggregate gains from AI may be modest and depend on how widely productivity improvements translate into meaningful economic outputs \citep{acemoglu2025simple}. Our findings provide complementary evidence from the domain of entrepreneurship: generative AI expands entry and these gains are concentrated in lower-commitment activity, and they may not translate proportionally into high-quality ventures. This highlights a key mechanism through which micro-level productivity improvements may fail to scale into aggregate innovation and macroeconomic outcomes.

\section{Empirical Setting}

We use Product Hunt as an empirical setting that offers a rare combination of scale and granularity for studying entrepreneurial organization in the presence of generative AI. Product Hunt is a prominent online platform where entrepreneurs publicly launch new technology products, including software products, mobile applications, web services and digital tools. Each launch is time-stamped and accompanied by detailed information on founding team composition, product category, product descriptions, subsequent updates, and user and platform-generated evaluations.

Product Hunt offers several advantages for studying entrepreneurial activity in the context of generative AI. First, the platform provides direct and observable measures of entrepreneurial entry which allows us to identify new ventures at the moment of public entry rather than relying on firm registrations, surveys or retrospective reporting. Second, Product Hunt records founding team size at launch, enabling analysis of founding structure at entry, a dimension that is difficult to observe in most large-scale datasets. Third, the platform generates rich, high-frequency outcome data, including user votes, daily rankings, which capture early market reception and relative visibility shortly after launch. Finally, the platform records post-launch updates and activity, allowing us to observe engagement and development behavior following entry.

For the main analyses, we examine monthly data from April 2020 to July 2025 and yearly data from 2019 to 2025 which provide balanced pre- and post-period around the release of ChatGPT-3.5 in November 2022. The sample includes 160,143 product launches.

\section{Data and Measures}

Our primary unit of analysis is a new product launch on Product Hunt, which we treat as an observable instance of entrepreneurial entry and venture creation. Consistent with the leading work in the entrepreneurship literature, venture creation is defined as the transition from a conceptual business idea to its first public market-facing realization rather than as a legal act of formal incorporation \citep{bhave1994process}.

\paragraph{Entrepreneurial Entry.}
Entry is measured as the number of new product launches. We aggregate product launches to the monthly level to construct panel datasets for regression analyses and to the annual level for complementary analyses. We distinguish between solo ventures (one maker) and team-based ventures (two or more makers), and further classify team-based ventures into size bins (2, 3, 4, 5, and 5+ members) to examine heterogeneity across founding structure. Founding team size is captured using the number of makers listed for each product at launch, as recorded by Product Hunt.

\paragraph{Product Category.}
Each entry is classified into broad categories based on keywords and tags associated on Product Hunt (e.g., Sales, Finance, and E-commerce). These categories reflect how products are positioned and described on the platform rather than mutually exclusive industry classifications. Because products may reference multiple category labels, a single product can be associated with more than one category. We use these categories to capture broad differences in the coordination and capability requirements associated with different types of products. To study cross-category heterogeneity, we compute each category’s pre-shock average share of solo ventures and classify categories as traditionally team-heavy or solo-heavy based on whether this share falls below or above the platform-wide mean.

\paragraph{One-shot Entry.}
We measure the persistence of newly launched ventures using an indicator for ``one-shot'' entry. Specifically, a venture is classified as one-shot if it exhibits no subsequent updates within 12 months of its initial launch. The outcome variable equals one if a venture does not release any follow-up update within 12 months, and zero otherwise. This measure captures whether ventures represent short-lived, low-commitment entry versus continued development following initial launch.

\paragraph{Product Distinctiveness.}
We measure the distinctiveness of newly launched products based on the rarity of their tag combinations relative to other products launched on the same day. Higher values indicate that a product reflects a more unique or less frequently observed combination of product attributes. This measure captures the extent to which ventures explore less familiar regions of the product space through new combinations of attributes.

\paragraph{Ranking-based Quality Outcome.}
We assess evaluation-based outcomes of ventures using Product Hunt’s daily rankings of launched products, generated by a platform-weighted algorithm that incorporates the number and velocity of user upvotes as well as user engagement. We compute the share of products appearing in different ranking tiers (e.g., Top10, Top30, Top50, and All), which allows us to distinguish between broad quality representation and concentration among the platform’s highest-ranked outcomes.

\section{Empirical Strategy}

Our research design leverages the public release of ChatGPT-3.5 in November 2022 as a shock that sharply expanded access to generative AI technology. The release marked a discrete and widely recognized shift in the availability and usability of large language models for both technical and non-technical users \citep{bick2024rapid}, making it well suited for studying changes in entrepreneurial entry and outcome before and after the introduction of generative AI.

We estimate a difference-in-differences (DiD) model with two-way fixed effects comparing monthly entry by solo founders and teams before and after the public release of ChatGPT-3.5. The outcome variable is the logarithm of the number of entries per month, separately for solo and team-founded ventures. The key coefficient captures differential change in entry rates between solo and team-based ventures after November 30, 2022. The specification controls for persistent differences between solo and team ventures and for month-specific shocks common to both groups. Then, event-study estimates allow us to assess whether pre-treatment trends were parallel and to trace the dynamic evolution of the effect. We estimate the model as follows:
\begingroup
\setlength{\abovedisplayskip}{6pt}
\setlength{\belowdisplayskip}{6pt}
\begin{equation}
\log(N_{gt}) = \alpha + \beta (Post_t \times Treat_g) + \gamma_g + \delta_t + \epsilon_{gt}
\label{eq:1}
\end{equation}
\begin{equation}
\log(N_{gt}) = \alpha + \sum_{k \neq 0} \beta_k (EventYear_{kt} \times Treat_g) + \gamma_g + \delta_t + \epsilon_{gt}
\label{eq:2}
\end{equation}
\endgroup

In both equations, $N_{gt}$ denotes the number of entrepreneurial entries in month $t$ by group $g$, where $g\in\{solo, team\}$. $Post_t$ equals one for months after November 30, 2022, and zero otherwise. $Treat_g$ equals one for solo entries and zero for team entries. $\gamma_g$ captures persistent differences between solo and team entry, and $\delta_t$ captures month-specific shocks common to both groups. $\epsilon_{gt}$ is an idiosyncratic error term. In \autoref{eq:2}, the coefficients $\beta_k$ trace this differential effect across event years, allowing us to assess pre-treatment trends and the dynamic evolution of the response. Identification relies on the assumption that absent the introduction of ChatGPT, solo and team entry would have followed parallel trends. Standard errors are clustered at the month level.

While the difference-in-differences analysis establishes how generative AI reshapes the level and composition of entrepreneurial entry, it does not reveal how these changes translate into differences in venture quality. To examine this, we next analyze how the distribution of quality outcomes varies by founding structure. We explore whether the introduction of ChatGPT-3.5 was associated with changes in the share of team-based ventures within each ranking tier. The outcome variable is an indicator equal to one if a newly launched venture is team-based, and zero otherwise. We estimate the following pre-post specification:
\begingroup
\setlength{\abovedisplayskip}{6pt}
\setlength{\belowdisplayskip}{6pt}
\begin{equation}
Team_{it}
=
\alpha
+
\beta Post_t
+
\sum_k \gamma_k Tier_{ik}
+
\sum_k \delta_k (Post_t \times Tier_{ik})
+
\epsilon_{it}
\label{eq:3}
\end{equation}
\endgroup

\noindent $Team_{it}$ equals one if venture $i$, launched on date $t$, is team-based, and zero otherwise. $Post_t$ equals one for launches after November 30, 2022, and zero otherwise. $Tier_{ik}$ denotes indicator variables for cumulative ranking tiers (Top 10, 30, 50, and All), with one tier serving as the reference category. $\epsilon_{it}$ is an idiosyncratic error term.

We complement the main analyses with additional regression-based and descriptive evidence on categorical heterogeneity in entry across product categories, one-shot entry, product distinctiveness, and team size distributions. The details of the analyses are reported in the \textit{Online Appendix}. Where appropriate, we employ difference-in-differences and event-study designs to examine changes over time and assess dynamic responses.

\section{Results}

We organize the results into four parts. First, we examine overall trends in entrepreneurial entry and decompose these trends into differences between solo and team ventures. Then, we assess how these changes vary across product categories. We next turn to the quality of the new entries: we start by examining the post-entry behavior and persistence, then evaluate team venture shares across ranking tiers and the distribution of team sizes among the top rank.

\subsection{Entrepreneurial Entry Following ChatGPT}

Entrepreneurial entry increased sharply following the release of ChatGPT-3.5, rising by 40\% from 29,326 in 2022 to 41,169 in 2023, and continued to grow toward twice its pre-release level by 2025 (summary statistics are reported in \autoref{tab:A1}). However, this expansion is not evenly distributed across founding structure.

\begin{figure}[H]
     \centering
     \caption{Entrepreneurial entry}     
     \begin{subfigure}[b]{0.495\linewidth}
         \centering
         \caption{Model-free}
         \label{fig:1a}
         \includegraphics[width=\linewidth]{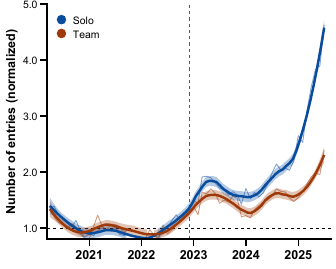} 
     \end{subfigure}
     \begin{subfigure}[b]{0.495\linewidth}
         \centering
         \caption{Flexible difference-in-differences}
         \label{fig:1b}
         \includegraphics[width=\linewidth]{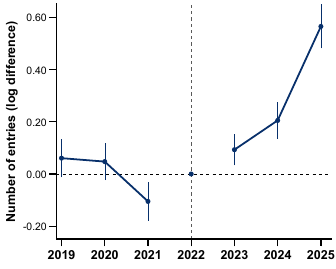}         
     \end{subfigure} \\
     \footnotesize \singlespacing \justifying \noindent \textit{Notes}. The vertical dashed line in both panels indicates the public release of ChatGPT on November 30, 2022. Each panel shows balanced pre- and post-treatment period, 32 months on each side for monthly series and 3 years on each side for yearly aggregation. (a) Monthly counts of entries by solo founders and teams, normalized to 1 in the pre-ChatGPT period. Thin lines show raw series, thicker lines summarize smoothed trends, and shaded bands indicate 95\% confidence intervals. (b) Flexible difference-in-differences estimates from \autoref{eq:2}, which regresses log monthly entry counts on indicators for event-year bins interacted with an indicator for solo entries, using 2022 (event year 0) as the omitted reference year. Each coefficient represents the difference in log entry counts in the indicated year relative to 2022. Points show coefficient estimates and vertical bars denote 95\% confidence intervals, with standard errors clustered by month.
\end{figure}

Decomposing the increase reveals that post-ChatGPT growth in entry is driven disproportionately by solo founders. Model-free evidence in \autoref{fig:1a} shows that while both solo and team entries increase following the shock, the growth in solo entries is substantially larger and the gap between solo and team  entries widens over time. \autoref{fig:1b} then illustrates the sharp growth in solo entries, relative to team entries, based on flexible difference-in-differences estimates. By 2025, solo entry grew nearly 90\% more than team entry ($\beta = 0.621$, $p < 0.001$). To assess robustness, we conduct placebo permutation tests in \autoref{fig:A2}, the results of which suggest that the effects are unlikely to arise from arbitrary or spurious patterns in the data or models. These patterns suggest that generative AI particularly enables solo entrepreneurship by lowering barriers to entry.

\subsection{Heterogeneity Across Product Categories}

\begin{figure}[H]
    \centering
    \caption{Changes in solo share across categories}     
    \begin{minipage}{0.7\linewidth}
        \centering
        \includegraphics[width=\linewidth]{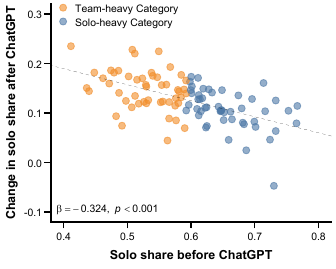}\\
        \footnotesize \singlespacing \justifying \noindent \textit{Notes}. Changes in solo entry share across categories from April 1, 2020 to July 31, 2025. The dashed line indicates the fitted linear regression of the change in solo share on pre-ChatGPT solo share. Reported $\beta$ coefficients and $p$-values are from this regression.
    \end{minipage}
    \label{fig:2}
\end{figure}

If generative AI disproportionately fueled solo ventures, where was this increase most pronounced?  We next examine whether the solo-driven rise in entry varied across product categories. \autoref{fig:2} plots each category’s post-shock solo-founder share against its pre-shock baseline and reveals a negative relationship. Categories that were more team-heavy before the shock experienced larger increases in solo entries, whereas categories that were solo-dominated showed more modest changes. This pattern is corroborated by a category-level regression, in which higher pre-shock solo-founder shares are associated with smaller post-shock increases in solo entry ($\beta = -0.324$, $p < 0.001$) \footnote {Because Product Hunt uses a non-hierarchical category tagging system, categories vary significantly in their level of abstraction and overlap. Our primary analysis uses the top 100 categories by launch frequency in the pre-shock period to capture the most representative market segments. In \autoref{fig:A3}, we demonstrate that our findings are robust to expanding or narrowing this scope (using the top 50, 150, and 200 categories), suggesting the observed trends are not artifacts of specific category definitions}.

\begin{figure}[H]
     \centering
     \caption{Changes in solo share across representative categories}
     \label{fig:3}
     \begin{subfigure}[b]{0.495\linewidth}
         \centering
         \caption{Team-heavy categories}
         \label{fig:3a}
         \includegraphics[width=\linewidth]{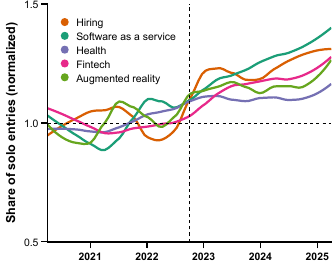}         
     \end{subfigure}
     \begin{subfigure}[b]{0.495\linewidth}
         \centering
         \caption{Solo-heavy categories}
         \label{fig:3b}
         \includegraphics[width=\linewidth]{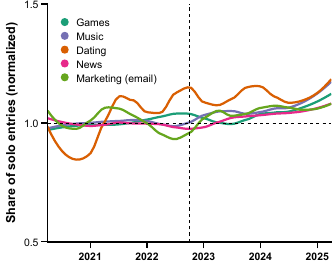}         
     \end{subfigure} \\
     \footnotesize \singlespacing \justifying \noindent \textit{Notes}. Quarterly share of solo entry across representative categories from April 1, 2020 to June 30, 2025. The vertical dashed line in all panels indicates the public release of ChatGPT-3.5 on November 30, 2022. All series are normalized to 1 in the pre-ChatGPT period.
\end{figure}

To provide further interpretability and assess the temporal dynamics underlying this relationship, \autoref{fig:3} presents quarterly time-series evidence for representative categories that are traditionally team-heavy and solo-heavy. We classify categories as team-heavy or solo-heavy depending on whether their pre-shock solo-founder share falls below or above the overall mean. Given that Product Hunt's category system reflects platform tagging conventions rather than formal industry classifications, we focus on categories that plausibly correspond to recognizable economic domains (e.g., Software-as-a-service and Fintech).

Consistent with the cross-sectional evidence, historically team-heavy categories displayed a pronounced and sustained increase in solo share following Q4 2022 (\autoref{fig:3a}), whereas historically solo-heavy categories exhibited comparatively modest changes (\autoref{fig:3b}). Importantly, pre-treatment trajectories appear relatively stable across both groups, supporting the interpretation that the observed divergence reflects a structural shift rather than short-term volatility. These patterns suggest that generative AI helped solo entrepreneurs enter domains where venture creation historically required collaboration among multiple human workers, at least partially substituting for the need for co-founders.

\subsection{Solo Entry and Experimentation}

\begin{figure}[H]
     \centering
     \caption{One-shot entry}
     \label{fig:4}
     \begin{subfigure}[b]{0.495\linewidth}
         \centering
         \caption{Solo}
         \label{fig:4a}
         \includegraphics[width=\linewidth]{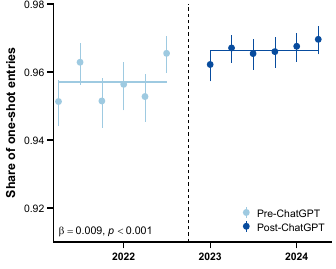}
     \end{subfigure}
     \begin{subfigure}[b]{0.495\linewidth}
         \centering
         \caption{Team}
         \label{fig:4b}
         \includegraphics[width=\linewidth]{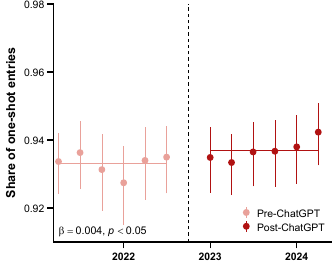}
     \end{subfigure}
     
     \footnotesize \singlespacing \justifying \noindent 
     \textit{Notes}. Quarterly shares of one-shot entries from 2020 Q2 to 2025 Q2. One-shot status is assessed over a 12-month window following entry which constrains observations at the beginning and end of the sample. The vertical dashed line indicates the release of ChatGPT-3.5 on November 30, 2022. Points show cohort-level one-shot shares with vertical whiskers denoting 95\% confidence intervals computed using the Wilson method. Horizontal segments indicate launch-weighted means in the balanced pre- and post-ChatGPT windows. Reported $\beta$ coefficients capture post-ChatGPT changes from regressions clustered by cohort.
\end{figure}

An important nuance that we highlight is the qualitative aspect of these solo ventures enabled by generative AI. We thus examine post-entry activities to check solo entry is associated with systematic changes in entrepreneurial behavior and persistence. We find that a substantial share of solo entries takes the form of \textit{one-shot entry}, in which an initial product launch is not followed by any subsequent update within twelve months \footnote{Given a mean time to first update of 9.6 months, most ventures that update do so within the first year. A 6-month window is overly restrictive, whereas a 24-month window captures extraneous long-tail re-entry. As such, 12-month window provides a conservative and meaningful measure of post-entry activity.}. \autoref{fig:4a} shows that the share of one-shot entries increased following the release of ChatGPT-3.5. The one-shot rate among solo ventures rises from 95.1\% to 97\% , more than twice the increase observed for team-based ventures in \autoref{fig:4b}.\footnote{Note that, because this is a tech product based online platform where anyone can post applications, the baseline one-shot rate is already relatively high at around 95\%. The fact that ChatGPT-3.5 further increased this figure to nearly 97\% represents a substantial effect.} The magnitude of the effect translates into approximately 215 additional such entries per year.

\begin{figure}[H]
     \centering
     \caption{Product distinctiveness}     
     \label{fig:5}
     
     \begin{subfigure}[b]{0.495\linewidth}
         \centering
         \caption{Model-free}
         \label{fig:5a}
         \includegraphics[width=\linewidth]{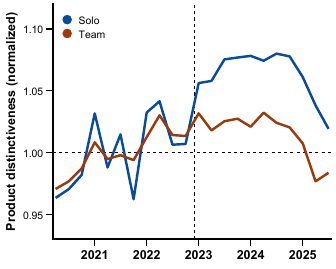} 
     \end{subfigure}
     \begin{subfigure}[b]{0.495\linewidth}
         \centering
         \caption{Flexible difference-in-differences}
         \label{fig:5b}
         \includegraphics[width=\linewidth]{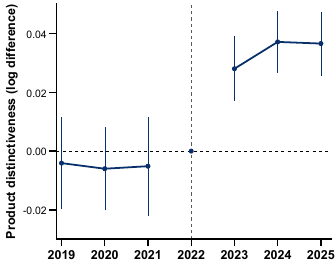}         
     \end{subfigure}
     
    \footnotesize \singlespacing \justifying \noindent 
    \textit{Notes}. The vertical dashed line in both panels indicates the public release of ChatGPT-3.5 on November 30, 2022. Each panel shows balanced pre- and post-treatment period, 10 quarters on each side for quarterly series and 3 years on each side for yearly aggregation. (a) Quarterly product distinctiveness by solo founders and teams, normalized to 1 in the pre-ChatGPT period. (b) Event-study estimates which regresses monthly mean product distinctiveness on event-year bins interacted with an indicator for solo entries, using 2022 as the omitted reference year. Distinctiveness is measured as the rarity of a product’s tag combination relative to all other products launched on the same day. Points show coefficient estimates and vertical bars denote 95\% confidence intervals, with standard errors clustered by month.
\end{figure}

To further assess whether this behavioral shift is accompanied by changes in the nature of ideas being tested, we examine the \textit{distinctiveness} of products introduced under this pattern of entry. Distinctiveness captures how rare a product's combination of attributes is relative to other products launched on the same day, with higher values indicating more unique combinations that are less frequently observed among peers. As shown in \autoref{fig:5a}, product distinctiveness increases following the release of ChatGPT-3.5 for both solo and team ventures, but the magnitude of change is substantially larger for solo founders. Consistently, the difference-in-differences estimates indicate that solo ventures experienced approximately 4\% larger increase in product distinctiveness relative to team ventures in the post-ChatGPT period ($\beta = 0.036$, $p < 0.001$; \autoref{fig:5b}, \autoref{tab:A4}). To assess robustness, we conduct placebo permutation tests in \autoref{fig:A4}, the results of which suggest that the effects are unlikely to arise from arbitrary or spurious patterns in the data or models.

The rise in one-shot entry suggests that AI-assisted entrepreneurial activity increasingly took the form of short-lived attempts to test ideas rather than sustained efforts to build and refine products. This is consistent with prior research showing that when uncertainty is high and commitment and coordination costs are low, individuals are more likely to initiate and abandon projects as part of a staged and reversible process \citep{trigeorgis2017real}. The increase in product distinctiveness further indicates that these efforts were not only more frequent, but also spanned a broader range of product attributes, reflecting an expansion of the idea space being explored. In this sense, generative AI enables solo entrepreneurs to engage in recombinant search. Prior work suggests that combining less familiar product attributes, while expanding the set of possibilities, tends to generate more variable and uncertain outcomes with lower average usefulness \citep{fleming2001recombinant}.  Together, these patterns point to broader, low-commitment experimentation for solo entries. Generative AI therefore does not simply lower the cost of entry, but reshapes entrepreneurial activity by enabling a more exploratory, trial-and-error approach to idea generation and testing, with important implications for quality outcomes.

\subsection{Team Entry and Quality Outcomes}

The key question, then, concerns which ventures produce the highest-quality products. Assessing the quality of new ventures at entry is inherently difficult, as early-stage ventures typically lack observable performance indicators such as revenue or user growth. Product Hunt provides a rare opportunity to observe early market evaluations at scale. The platform features a sophisticated daily ranking system in which newly launched products are assessed through community votes and engagement from developers, investors, and domain experts. These rankings aggregate dispersed judgments about novelty and perceived potential at the moment of entry, offering a useful proxy for early-stage venture quality.

\begin{figure}[H]
     \centering
     \caption{Team representation in top ranks}     
     \label{fig:event_pcounaat}
     \begin{subfigure}[b]{0.495\linewidth}
         \centering
         \caption{Team share across ranking}
         \label{fig:6a}
         \includegraphics[width=\linewidth]{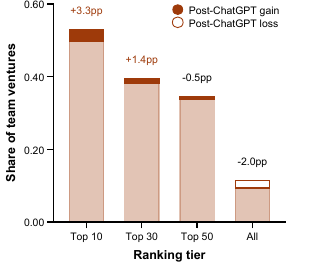}         
     \end{subfigure}
     \begin{subfigure}[b]{0.495\linewidth}
         \centering
         \caption{Top 10 share by team size}
         \label{fig:6b}
         \includegraphics[width=\linewidth]{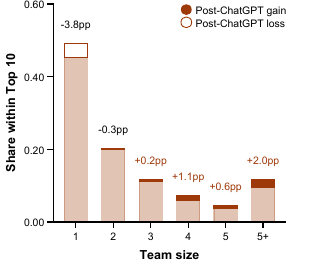}         
     \end{subfigure} \\
     \footnotesize \singlespacing \justifying \noindent \textit{Notes}. Share of team-founded ventures across platform ranking tiers before and after the public release of ChatGPT on November 30, 2022. The pre-ChatGPT period is defined as April 1, 2020 to November 30, 2022, and the post-ChatGPT period as December 1, 2022 to July 31, 2025. (a) Change in the share of team-founded ventures across cumulative ranking tiers (Top 10, Top 30, Top 50, and all ventures). (b) Change in the share of team-founded ventures within the Top 10, disaggregated by team size.
\end{figure}

Despite the surge in solo entries, team-based ventures continue to dominate the top tiers of platform's daily rankings. \autoref{fig:6a} shows that teams account for the majority of the Top 10, with their share even increasing from 50\% to 53\% ($p = 0.006$) following the release of ChatGPT-3.5. A similar pattern holds for the Top 30, where team representation rises from 38\% to 40\% ($p = 0.057$).  As the definition of ``top'' is broadened to include relatively lower-quality products, however, team representation declines and the post-ChatGPT gain reverses. Across all products, the share of team-based ventures decreased by two percentage points. That is, while generative AI enables entry, much of the expansion in solo ventures does not translate proportionally into the highest-quality outcomes. Corresponding regression estimates are reported in \autoref{tab:A6} and the longitudinal evolution of these patterns is further illustrated in \autoref{fig:A5}. 

To further examine the quality effect by team size, we focus on Top 10 and disaggregate the share by team size (\autoref{fig:6b}). Post-ChatGPT gains are concentrated among larger teams, particularly those with more than three members; the extent of increase was greatest for teams with five or more members. By contrast, solo and duo ventures experience declines relative to the pre-ChatGPT baseline.

This supports the idea that teams producing the highest-quality products benefit from the diverse knowledge and experience of human co-founders, which may not be readily substituted by inputs from a single or even multiple AI models. In particular, while generative AI expands access to standardized and codified capabilities, producing high-quality and differentiated outcomes continues to depend on integrating tacit and experiential knowledge that is difficult to fully encode \citep{grant1996knowledge,kogut1992knowledge}. Such knowledge is often distributed across individuals and becomes valuable through recombination and collective judgment. Generative AI therefore does not fully substitute for teams in producing top-tier outcomes. This point is especially salient given evidence that different AI models often generate highly similar responses to open-ended questions, exhibiting an artificial hivemind \citep{jiang2025artificial}.

Together, these results point to an important asymmetry in how generative AI reshapes entrepreneurship. While AI broadly facilitates entry, it does not appear to relax the organizational requirements for reaching the top of the quality distribution. The findings are consistent with limits to the substitutability of generative AI for context-dependent judgment and experiential know-how, particularly in environments characterized by evolving goals and complex demands \citep{acemoglu2025simple}. In an AI-enabled setting marked by widespread experimentation, superior outcomes at the upper tail appear increasingly tied to the ability to combine complementary skills and sustain development beyond initial launch, rather than to entry alone.

\section{Conclusion}

Generative AI reshapes entrepreneurial activity by lowering barriers to entry and enabling widespread experimentation, while the organizational advantages of team-based ventures in producing top-quality outcomes persist. These differential effects point to a growing divergence between the quantity and quality of entrepreneurial activity that is likely to shape how technological change translates into economic impact.

This divergence has important implications for how entrepreneurship is evaluated and supported. Policies that reduce frictions to entry can broaden participation and encourage experimentation, but do not automatically generate high-quality outcomes. Achieving top-tier performance appears to continue to depend on the recombination of diverse human knowledge and experience.

More broadly, generative AI reallocates, rather than eliminates, constraints in venture creation. It shifts bottlenecks away from early-stage development and toward longer-term development and improvement. In an AI-enabled economy, the central challenge is therefore not simply expanding entry, but organizing human expertise effectively alongside AI to convert widespread experimentation into sustained economic value.

It is also noteworthy that, while generative tools expand the range of ideas explored, as reflected in the increase in product distinctiveness particularly among solo ventures, these gains in distinctiveness at entry fade away gradually. Over time, the use of similar generative tools and shared inputs may draw entrepreneurs toward similar solution spaces which could reduce the collective novelty \citep{doshi2024generative, jiang2025artificial}. Similarly, early public experimentation can accelerate learning but it may also expose ventures to imitation in digital environments where replication is rapid and formal protection is limited \citep{contigiani2023experiment}. Expanded experimentation may therefore intensify imitation pressures and compress returns to sustained development, heightening the importance of institutional clarity around appropriability and support for organizational capability building.

We acknowledge that this study examines entrepreneurial activity within a single digital platform focused on software product launches, which may not fully capture venture creation in other sectors or later-stage business outcomes. Accordingly, generalization should be made with caution. Future research could examine whether similar patterns emerge in settings involving incorporated firms, capital-intensive industries, or longer-term measures of performance such as survival, growth, and innovation.

\clearpage

\bibliographystyle{apalike}
\begin{singlespace}
  \begin{small}
    \bibliography{producthunt}
  \end{small}
\end{singlespace}

\csname efloat@restorefloats\endcsname
\setcounter{table}{0}
\renewcommand{\thetable}{\Alph{section}\arabic{table}}
\renewcommand{\thefigure}{\thesection\arabic{figure}}

\newpage
\appendix

\renewcommand{\thesection}{Appendix \Alph{section}}
\renewcommand{\thesubsection}{\Alph{section}\arabic{subsection}}

\setcounter{section}{0}
\setcounter{subsection}{0}

\renewcommand{\thefigure}{A\arabic{figure}}
\setcounter{figure}{0}

\renewcommand{\thetable}{A\arabic{table}}
\setcounter{table}{0}

\section{Product Hunt Platform Background}

\paragraph*{Product Hunt Overview}
Product Hunt is a U.S.-based online platform launched in 2013 that serves as a marketplace for discovering and evaluating newly launched digital products. As of July 30, 2025, more than 315,000 products had been launched on the platform. Often described as a go-to destination for entrepreneurs, investors, product managers, developers and technology enthusiasts, Product Hunt began within the Silicon Valley ecosystem but now attracts global participation. The platform enables users to discover, test and evaluate new products and engage in discussion through comments and community forums. These products are ranked daily, creating a visible and time-bound competitive environment. For entrepreneurs, Product Hunt represents a point at which their idea transitions into a publicly visible venture, facing open market exposure. In many cases, this stage precedes formal incorporation, capturing an early phase of venture emergence rather than firm creation. The platform offers immediate visibility and feedback from a broad audience, shaping early perceptions of demand and quality.

\paragraph*{Product Discovery}
When users arrive at the Product Hunt website, they encounter multiple pathways for discovering newly launched products. They may directly search for specific offerings, browse curated listings such as “Top Products Launching Today,” or explore products by category. Most visits expose users to recently launched products. However, what users seek to discover varies across audiences. Non-technical users may browse to stay informed about broader technology trends while technology enthusiasts often look for novel or emerging tools to try. Entrepreneurs may explore the platform for inspiration or competitive awareness and investors use it to identify new products and ventures at their earliest stages. 

\paragraph*{Product Launch and Evaluation}
Entrepreneurs introduce their product by providing descriptive and visual information including product summaries, feature explanations, screenshots, and demonstration materials. Launch pages typically include a direct link to the product’s website or application store, enabling users to access or try the offering. Each product page also displays information about the makers, such as founders, designers, or engineers associated with the launch, allowing users to observe the individuals or team behind the venture. Users evaluate launched products through upvotes and public comments.

\paragraph*{Subsequent Launches (updates)}
A first product launch marks the initial introduction of a product to the platform. As ventures evolve, they may release subsequent updates to the same product, which are recorded under the original product page and same product code in the data. These updates can range from minor improvements such as bug fixes to more substantial changes including the introduction of new features, interface redesigns or expanded functionality. Regardless of scope, each update undergoes the same ranking and evaluation process as an initial launch, receiving upvotes, comments, and daily ranking placement.

\paragraph*{Daily Ranking}
Product Hunt assigns each launch a point score that determines its placement in the daily rankings. Points are calculated based on authentic user engagement rather than raw upvote counts alone. In addition to upvotes, the ranking system considers other engagement signals, including comments and community participation. As a result, rankings reflect a weighted measure of overall market traction rather than simple vote totals.

\clearpage

\section{Supplementary Methods}
Below, we provide additional methodological details for the supplementary analyses used in the study.

\subsection{Changes in solo entry share across categories}
We examine whether the increase in solo entrepreneurship following the release of ChatGPT varies systematically across categories depending on their pre-period founding structure. Specifically, we test whether categories that were historically more team-oriented experienced larger post-ChatGPT increases in solo entry. For each category c, we compute the average solo share before and after November 30, 2022. The outcome variable is the change in solo share between the post- and pre-periods. We estimate the following cross-category regression:
\renewcommand{\theequation}{S\arabic{equation}}
\setcounter{equation}{0}
\begin{equation}
\Delta \text{SoloShare}_c = \alpha + \beta \text{PreShare}_c + \epsilon_c
\label{eq:S1}
\end{equation}

\( \Delta SoloShare_c \) denotes the change in the average monthly solo share for category \( c \) between the post- and pre-ChatGPT periods. \( PreShare_c \) denotes the average monthly solo share for category \( c \) during the pre-ChatGPT period. \( \epsilon_c \) is an idiosyncratic error term. The coefficient \( \beta \) in captures whether categories with lower pre-period solo shares (i.e., more team-oriented categories) experienced larger increases in solo entry following the introduction of generative AI. A negative coefficient indicates that historically team-oriented categories exhibit larger post-ChatGPT increases in solo entrepreneurship. 

\begin{figure}[!htbp]
     \centering
     \caption{Changes in solo share across top-n categories}
     \label{fig:A3}
     \begin{subfigure}[b]{0.495\linewidth}
         \centering
         \caption{Top 50 categories}
         \label{fig:A3a}
         \includegraphics[width=\linewidth]{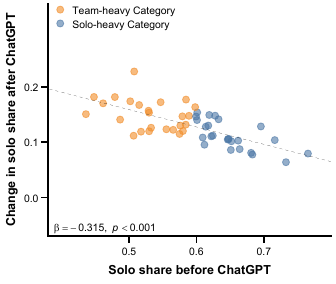}             
     \end{subfigure}
     \begin{subfigure}[b]{0.495\linewidth}
         \centering
         \caption{Top 100 categories}
         \label{fig:A3b}
         \includegraphics[width=\linewidth]{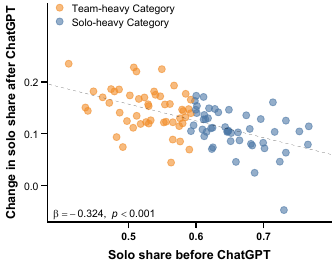}
     \end{subfigure} 
     \begin{subfigure}[b]{0.495\linewidth}
         \centering
         \caption{Top 150 categories}
         \label{fig:A3c}
         \includegraphics[width=\linewidth]{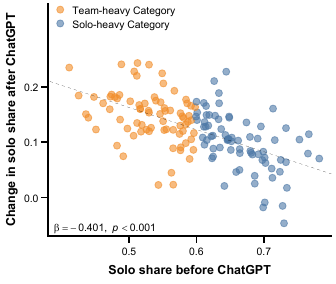}
     \end{subfigure} 
     \begin{subfigure}[b]{0.495\linewidth}
         \centering
         \caption{Top 200 categories}
         \label{fig:A3d}
         \includegraphics[width=\linewidth]{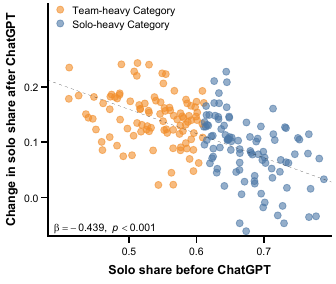}
     \end{subfigure} \\
     \footnotesize \justifying \noindent \textit{Notes}. Changes in solo entry share across categories from April 1, 2020 to July 31, 2025, using alternative category cutoffs to complement the top 100 specification used in the main analysis. Each panel plots, for categories ranked by pre-ChatGPT launch frequency, the average solo share before the public release of ChatGPT on November 30, 2022 against the change in solo share after ChatGPT. Panels show (a) Top 50 categories, (b) Top 100 categories, (c) Top 150 categories, and (d) Top 200 categories. Each point represents a category. Categories are classified as team-heavy or solo-heavy based on whether their pre-period solo share falls below or above the median across categories. The dashed line indicates the fitted linear regression of the change in solo share on pre-ChatGPT solo share. Reported $\beta$ coefficients and corresponding $p$-values are from category-level regressions.
\end{figure}

We complement the main specification, which focuses on the top 100 most frequent pre-ChatGPT categories, by replicating the analysis using alternative sets of top-\( n \) most frequent categories (50, 150, and 200), reported in \autoref{fig:A3}.

\subsection{Share of one-shot entries}
We examine whether the introduction of ChatGPT was associated with a change in the persistence of newly launched ventures. Specifically, we test whether ventures launched after November 30, 2022 were more likely to be “one-shot” entries, defined as ventures that exhibit no subsequent updates within 12 months of their initial launch. The outcome variable is an indicator equal to one if a venture does not update within 12 months of entry, and zero otherwise. We estimate the following pre-post specification:
\begin{equation}
\text{OneShot}_{iq} = \alpha + \beta \text{Post}_q + \varepsilon_{iq}
\label{eq:S2}
\end{equation}

\( OneShot_{iq} \) equals one if product \( i \), entering in quarter \( q \), has no follow-up post within 12 months, and zero otherwise. \( Post_q \) equals one for entry quarters after November 30, 2022, and zero otherwise. \( \varepsilon_{iq} \) is an idiosyncratic error term. The coefficient \( \beta \) in captures whether the probability of one-shot entry changed following the public release of ChatGPT. Standard errors are clustered at the entry-quarter level. 

\subsection{Product distinctiveness}

We examine whether the introduction of ChatGPT was associated with changes in the distinctiveness of newly launched products. Specifically, we test whether products developed by solo founders exhibit differential changes in distinctiveness relative to team-based ventures following November 30, 2022. Distinctiveness is measured as the rarity of a product’s tag combination relative to other products launched on the same day, with higher values indicating more unique or less frequently observed combinations.

We estimate a difference-in-differences specification of the form:
\begin{equation}
D_{gt} = \alpha + \beta (\text{Post}_t \times \text{Solo}_g) + \gamma_t + \delta_g + \varepsilon_{gt}
\label{eq:S3}
\end{equation}

where \( D_{gt} \) denotes the mean product distinctiveness for founder type \( g \in \{\text{Solo, Team}\} \) in month \( t \). \( \text{Post}_t \) is an indicator equal to one for months following November 30, 2022, and zero otherwise. \( \text{Solo}_g \) is an indicator for solo-founded ventures. \( \gamma_t \) represents month fixed effects and \( \delta_g \) denotes founding structure fixed effects. The coefficient \( \beta \) captures the differential change in product distinctiveness for solo ventures relative to team ventures in the post-ChatGPT period. Standard errors are clustered at the month level.

To examine the dynamics of this relationship, we complement the baseline specification with an event-study design that replaces the post indicator with a series of event-year indicators relative to 2022:
\begin{equation}
D_{gt} = \alpha + \sum_{k \neq 0} \beta_k \left( \text{EventYear}_k \times \text{Solo}_g \right) + \gamma_t + \delta_g + \varepsilon_{gt}
\label{eq:S4}
\end{equation}

where \( \text{EventYear}_k \) indexes calendar years relative to 2022 (event year 0). The coefficients \( \beta_k \) trace the evolution of the distinctiveness gap between solo and team ventures over time.

The baseline difference-in-differences estimates are reported in \autoref{tab:A4}. Event-study estimates are reported in \autoref{tab:A5}.

\subsection{Share of team-founded ventures within the Top 10}
We examine whether the introduction of ChatGPT was associated with changes in the distribution of team sizes among ventures ranked in the top 10. The outcome variable is an indicator equal to one if a newly launched venture belongs to a given team size category (1, 2, 3, 4, 5, or 5+ founders), and zero otherwise. We estimate the following pre-post specification:
\begin{equation}
y_{ist} = \alpha + \beta \text{Post}_t + \gamma_s \text{Size}_{is} + \delta_s (\text{Post}_t \times \text{Size}_{is}) + \varepsilon_{ist}
\label{eq:S5}
\end{equation}

\( y_{ist} \) equals one if venture \( i \), launched at time \( t \), belongs to team-size category \( s \), and zero otherwise. \( Post_t \) equals one for launches after November 30, 2022, and zero otherwise. \( Size_{is} \) denotes indicator variables for team sizes 2, 3, 4, 5, and 5+, with solo ventures (team size 1) serving as the reference category. \( \varepsilon_{ist} \) is an idiosyncratic error term. The coefficient \( \beta \) in captures the post-ChatGPT change in the share of solo ventures, while \( \beta + \delta_s \) captures the corresponding post-ChatGPT change for team size \( s \). Standard errors are clustered at the launch-date level. 

\begin{figure}[!htbp]
     \centering
     \caption{Changes in share across team size within Top 10}
     \label{fig:A6}
    \begin{minipage}{0.7\linewidth}
         \centering
         \includegraphics[width=\linewidth]{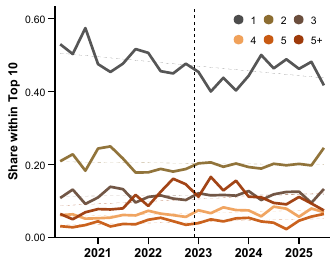}\\
         \footnotesize \justifying \noindent \textit{Notes}. Quarterly share of ventures by founding team size within Top 10 from April 1, 2020 to June 30, 2025. The vertical dashed line in all panels indicates the public release of ChatGPT on November 30, 2022. Shares are calculated relative to the total number of Top 10 launches in each quarter. Sloped dashed lines represent fitted linear trends for each team size.
    \end{minipage}
\end{figure}

Regression results are reported in \autoref{tab:A7}. We complement this analysis with quarterly time-series plots that display team-size shares within the top 10, including 95\% confidence intervals and launch-weighted pre- and post-ChatGPT averages, in \autoref{fig:A6}.

\clearpage

\section{Supplementary Figures and Tables}

\begin{figure}[!htbp]
     \centering
     \caption{Entrepreneurial entry by team size}     
     \label{fig:A1}
     \begin{subfigure}[b]{0.495\linewidth}
         \centering
         \caption{Normalized trend by team size}
         \label{fig:A1a}
         \includegraphics[width=\linewidth]{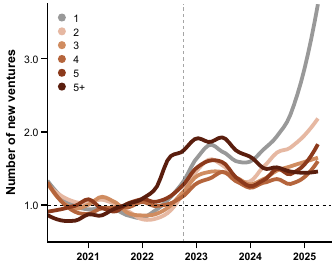}         
     \end{subfigure}
     \begin{subfigure}[b]{0.495\linewidth}
         \centering
         \caption{Share change by team size}
         \label{fig:A1b}
         \includegraphics[width=\linewidth]{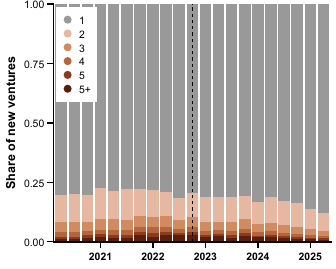}
     \end{subfigure} \\
     \footnotesize \justifying \noindent \textit{Notes}. Quarterly entrepreneurial entry from April 1, 2020 to June 30, 2025, disaggregated by team size. The vertical dashed line in all panels indicates the public release of ChatGPT on November 30, 2022. (a) Monthly counts of new ventures launched by team size, normalized to 1 in the pre-ChatGPT period. (b) Share of new ventures launched by team size.
\end{figure}

\clearpage

\begin{figure}[!htbp]
     \centering
     \caption{Placebo permutation tests for entrepreneurial entry}
     \label{fig:A2}
    \begin{minipage}{0.7\linewidth}
         \centering
         \includegraphics[width=\linewidth]{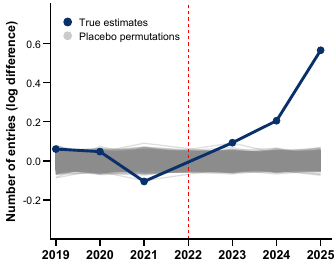}\\
         \footnotesize \justifying \noindent \textit{Notes}. Yearly event-study estimates with the distribution of 1,000 placebo permutation estimates. The dependent variable is the log-transformed number of entrepreneurial entries. Blue lines and circle markers represent the true event-study estimates whereas gray lines represent placebo permutations. In the placebo tests, solo versus team founder labels are randomly reassigned across entrepreneurial entries while preserving the overall sample composition. The vertical dashed line indicates the public release of ChatGPT-3.5 in 2022.
    \end{minipage}
\end{figure}

\clearpage

\clearpage

\begin{figure}[!htbp]
     \centering
     \caption{Placebo permutation tests for product distinctiveness}
     \label{fig:A4}
    \begin{minipage}{0.7\linewidth}
         \centering
         \includegraphics[width=\linewidth]{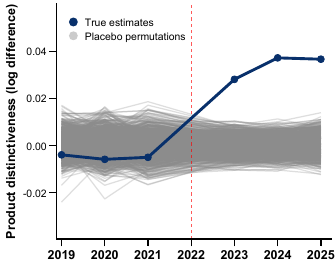}\\
         \footnotesize \justifying \noindent \textit{Notes}. Yearly event-study estimates with the distribution of 1,000 placebo permutation estimates. The dependent variable is the log-transformed product distinctiveness. Blue lines and circle markers represent the true event-study estimates whereas gray lines represent placebo permutations. In the placebo tests, solo versus team founder labels are randomly reassigned within month while preserving the monthly composition of entrepreneurial entries. The vertical dashed line indicates the public release of ChatGPT-3.5 in 2022.
    \end{minipage}
\end{figure}

\clearpage

\begin{figure}[!htbp]
    \centering
    \caption{Changes in team share across rank tiers}
    \label{fig:A5}
    \begin{minipage}{0.7\linewidth}
        \centering
        \includegraphics[width=\linewidth]{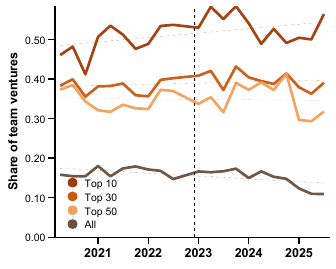}
        \footnotesize
        \justifying
        \noindent \textit{Notes}. Quarterly share of team ventures across rank tiers from April 1, 2020 to June 30, 2025. The vertical dashed line indicates the public release of ChatGPT on November 30, 2022. Rank tiers correspond to Top 10, Top 30, Top 50, and All ventures. Shares are calculated relative to the total number of launches within each tier and quarter. Sloped dashed lines represent fitted linear time trends for each rank tier.
    \end{minipage}
\end{figure}

\clearpage

\clearpage

\begin{table}[!htbp]
    \centering
    \begin{threeparttable}
    \caption{Entrepreneurial entry descriptive statistics}
    \label{tab:A1}
    \begin{tabular}{lrrrrrrrr}
    \toprule
    & \multicolumn{4}{c}{Pre-ChatGPT} & \multicolumn{4}{c}{Post-ChatGPT} \\
    \cmidrule(lr){2-5} \cmidrule(lr){6-9}
    Variables & Mean & SD & Min & Max & Mean & SD & Min & Max \\
    \midrule
    Total ventures & 1,664.53 & 260.26 & 1,298 & 2,254 & 3,339.94 & 1,152.12 & 2,202 & 6,934 \\
    Solo ventures  & 1,318.78 & 220.65 & 1,011 & 1,837 & 2,798.44 & 1,073.05 & 1,750 & 6,106 \\
    Team ventures  & 345.75   & 49.60  & 262   & 481   & 541.50   & 87.73    & 409   & 828 \\
    \bottomrule
    \end{tabular}
    \begin{tablenotes}[flushleft]
    \footnotesize
    \item[] \textit{Notes}. Monthly descriptive statistics for new venture creations on Product Hunt from April 1, 2020 to July 31, 2025. Statistics are reported separately for the pre-ChatGPT period (April 2020 to November 2022) and the post-ChatGPT period (December 2022 to July 2025).
    \end{tablenotes}
    \end{threeparttable}
\end{table}

\clearpage

\begin{table}[!htbp]
    \centering
    \begin{threeparttable}
    \caption{Solo vs. team entry difference-in-differences estimate}
    \label{tab:A2}
    \begin{tabular}{l S[table-format=1.3]}
    \toprule
     & {Entrepreneurial entry: solo vs. team} \\
    \midrule
    Post x Treat & 0.264*** \\
     & {(0.040)} \\
    Observations & {128} \\
    Founding structure FE & {YES} \\
    Time FE & {YES} \\
    \bottomrule
    \end{tabular} 
    \begin{tablenotes}[flushleft]
    \footnotesize \item[] \textit{Notes}. Difference-in-differences specifications based on \autoref{eq:1}, examining changes in monthly venture creation following the public release of ChatGPT in November 2022. The dependent variable is the logarithm of one plus the number of new venture creations per month by founding structure (solo vs. team). Standard errors are clustered at the month level. The unit of observation is founding structure by month. Robust standard errors are in brackets; ***$p < 0.001$, **$p < 0.01$, *$p < 0.05$.
    \end{tablenotes}
    \end{threeparttable}
\end{table}

\clearpage

\FloatBarrier
\begin{table}[!htbp]
    \centering
    \begin{threeparttable}
    \caption{Solo vs.\ team entry event-study estimates}
    \label{tab:A3}
    \begin{tabular}{crrr}
    \toprule
    Event year & Estimate & SE & p-value \\
    \midrule
    $-3$ & 0.061 & 0.037 & 0.108 \\
    $-2$ & 0.047 & 0.036 & 0.192 \\
    $-1$ & -0.105 & 0.037 & 0.006 \\
    1 & 0.093 & 0.031 & 0.003 \\
    2 & 0.205 & 0.036 & $<0.001$ \\
    3 & 0.621 & 0.046 & $<0.001$ \\
    \bottomrule
    \end{tabular}
    \begin{tablenotes}[flushleft]
    \footnotesize
    \item[] \textit{Notes}. Event-study specifications based on \autoref{eq:2}, examining changes in monthly entrepreneurial entry following the public release of ChatGPT in November 2022. The dependent variable is the logarithm of one plus the number of new venture creations per month by founding structure (solo vs.\ team). Event year indexes calendar years relative to the release, with coefficients reported relative to calendar year 2022 (event year 0). Standard errors are clustered at the month level. The unit of observation is founding structure by month. \\ \\
    The event-study estimates indicate that earlier pre-treatment periods (event years -3 and -2, 2019 and 2020) exhibit coefficients that are small and statistically indistinguishable from zero which is consistent with parallel trends. In contrast, the estimate for the year immediately preceding the release (event year -1, 2021) is negative and statistically significant (-0.105, $p = 0.006$). Further analysis shows that this result is driven by localized fluctuations in a small number of months (specifically February, March, and August 2021) which are temporally distant from the November 2022 treatment. Excluding these three months attenuates the estimate to -0.069 ($p = 0.053$) making it statistically insignificant. This pattern suggests that the apparent pre-treatment difference in one of the three pre-trend years reflects temporary, localized variation rather than a sustained divergence in trends. This pattern is also evident in the monthly trends shown in \autoref{fig:1a} in the main text.
    \end{tablenotes}
    \end{threeparttable}
\end{table}

\clearpage

\FloatBarrier
\begin{table}[!htbp]
    \centering
    \begin{threeparttable}
    \caption{Solo vs.\ team product distinctiveness difference-in-differences estimate}
    \label{tab:A4}
    \begin{tabular}{l S[table-format=1.3]}
    \toprule
     & {Product distinctiveness: Solo vs.\ Team} \\
    \midrule
    Post x Treat & 0.036*** \\
     & {(0.004)} \\
    Observations & {128} \\
    Founding structure FE & {YES} \\
    Time FE & {YES} \\
    \bottomrule
    \end{tabular}
    \begin{tablenotes}[flushleft]
    \footnotesize
    \item[] \textit{Notes}. Difference-in-differences specification examining changes in monthly product distinctiveness following the public release of ChatGPT in November 2022. The dependent variable is the monthly mean product distinctiveness by founding structure (solo vs.\ team). Product distinctiveness is measured as the negative log of the number of products sharing the same tag combination on a given day, such that higher values indicate more distinctive product attributes. Standard errors are clustered at the month level. The unit of observation is founding structure by month. Robust standard errors are in brackets; ***$p < 0.001$, **$p < 0.01$, *$p < 0.05$.
    \end{tablenotes}
    \end{threeparttable}
\end{table}

\clearpage

\FloatBarrier
\begin{table}[!htbp]
    \centering
    \begin{threeparttable}
    \caption{Solo vs.\ team product distinctiveness event-study estimates}
    \label{tab:A5}
    \begin{tabular}{crrr}
    \toprule
    Event year & Estimate & SE & p-value \\
    \midrule
    $-3$ & -0.004 & 0.008 & 0.613 \\
    $-2$ & -0.006 & 0.007 & 0.411 \\
    $-1$ & -0.005 & 0.009 & 0.552 \\
    1 & 0.028 & 0.006 & $<0.001$ \\
    2 & 0.037 & 0.005 & $<0.001$ \\
    3 & 0.037 & 0.006 & $<0.001$ \\
    \bottomrule
    \end{tabular}
    \begin{tablenotes}[flushleft]
    \footnotesize
    \item[] \textit{Notes}. Event-study specification examining changes in monthly product distinctiveness following the public release of ChatGPT in November 2022. The dependent variable is the monthly mean product distinctiveness by founding structure (solo vs.\ team). Product distinctiveness is measured as the negative log of the number of products sharing the same tag combination on a given day, such that higher values indicate more distinctive product positioning. Event year indexes calendar years relative to the release, with coefficients reported relative to calendar year 2022 (event year 0). Standard errors are clustered at the month level and reported in parentheses. The unit of observation is founding structure by month. \\ \\
    The event-study estimates indicate that pre-treatment coefficients are small and statistically indistinguishable from zero, providing no evidence of differential pre-trends between solo and team entries prior to the release of ChatGPT. The absence of systematic differences in the pre-period supports the validity of the identification strategy and suggests that the post-treatment divergence reflects a structural shift rather than a continuation of prior trends.
    \end{tablenotes}
    \end{threeparttable}
\end{table}

\clearpage

\begin{table}[!htbp]
    \centering
    \begin{threeparttable}
    \caption{Changes in team share across ranking tiers}
    \label{tab:A6}
    \begin{tabular}{lrrrrrrr}
    \toprule
    Tier & Pre share & Post share & Change & 95\% CI & p-value & N pre & N post \\
    \midrule
    Top 10 & 0.497 & 0.530 & 0.033 & [0.009, 0.056] & 0.006 & 6,616 & 6,919 \\
    Top 30 & 0.381 & 0.395 & 0.014 & [0.000, 0.029] & 0.057 & 18,043 & 18,867 \\
    Top 50 & 0.344 & 0.339 & -0.005 & [-0.018, 0.007] & 0.389 & 22,196 & 26,831 \\
    All & 0.342 & 0.191 & -0.151 & [-0.164, -0.139] & $<0.001$ & 22,405 & 58,353 \\
    \bottomrule
    \end{tabular}

    \begin{tablenotes}[flushleft]
    \footnotesize
    \item[] \textit{Notes}. The table reports estimated pre- and post-ChatGPT changes in the share of team-based ventures within each ranking tier. The pre-period spans April 2020 to November 2022, and the post-period spans December 2022 to July 2025. Change is defined as the post minus pre difference. Estimates are obtained from the linear probability model in Equation 3, which interacts the post indicator with tier indicators. Confidence intervals are 95\% and are based on standard errors clustered by launch date.
    \end{tablenotes}
    \end{threeparttable}
\end{table}

\clearpage

\FloatBarrier
\begin{table}[!htbp]
    \centering
    \begin{threeparttable}
    \caption{Changes in distribution of team sizes within Top 10 rankings}
    \label{tab:A7}
    \begin{tabular}{lrrrrrrr}
    \toprule
    Team size & Pre share & Post share & Change & 95\% CI & p-value & N pre & N post \\
    \midrule
    1 & 0.491 & 0.453 & -0.038 & [-0.061, -0.014] & 0.002 & 3,171 & 3,036 \\
    2 & 0.202 & 0.200 & -0.003 & [-0.017, 0.011] & 0.723 & 1,308 & 1,340 \\
    3 & 0.113 & 0.115 & 0.002 & [-0.009, 0.014] & 0.676 & 729 & 773 \\
    4 & 0.060 & 0.071 & 0.011 & [0.003, 0.020] & 0.010 & 388 & 479 \\
    5 & 0.038 & 0.044 & 0.006 & [-0.001, 0.013] & 0.087 & 247 & 297 \\
    5+ & 0.096 & 0.116 & 0.020 & [0.008, 0.033] & 0.001 & 618 & 778 \\
    \bottomrule
    \end{tabular}

    \begin{tablenotes}[flushleft]
    \footnotesize
    \item[] \textit{Notes}. The table reports estimated pre- and post-ChatGPT changes in the share of launches by team size among products ranked in the Top 10. The pre-period spans April 2020 to November 2022, and the post-period spans December 2022 to July 2025. Change is defined as the post minus pre difference. Estimates are obtained from linear probability models, estimated separately for each team size, in which the dependent variable is an indicator for the given team size and the post indicator captures the effect of ChatGPT's release. Confidence intervals are 95\% and are based on standard errors clustered by launch date.
    \end{tablenotes}
    \end{threeparttable}
\end{table}

\end{document}